\documentclass[dvips,12pt]{article}
\usepackage{amsmath,amssymb,exscale}   
\usepackage{array,multicol}
\usepackage{afterpage,float,flafter}   
\usepackage{epsfig,rotating,pifont,fancybox}   
\makeatletter   
\@addtoreset{equation}{section}   
\makeatother   
   
\def\simlt{\stackrel{<}{{}_\sim}}
\def\simgt{\stackrel{>}{{}_\sim}}
\newcommand{\bea}{\begin{eqnarray}}   
\newcommand{\eea}{\end{eqnarray}}   
   
\newcommand{\NPB}[3]{\emph{ Nucl.~Phys.} \textbf{B#1} (#2) #3}   
\newcommand{\PLB}[3]{\emph{ Phys.~Lett.} \textbf{B#1} (#2) #3}   
\newcommand{\PRD}[3]{\emph{ Phys.~Rev.} \textbf{D#1} (#2) #3}   
\newcommand{\PRL}[3]{\emph{ Phys.~Rev.~Lett.} \textbf{#1} (#2) #3}

\newcommand{\MPL}[3]{\emph{ Mod.~Phys.~Lett.} \textbf{A#1} (#2) #3}   
\newcommand{\PR}[3]{\emph{ Phys.~Rep.} \textbf{#1} (#2) #3}   
\newcommand{\RMP}[3]{\emph{ Rev.~Mod.~Phys.} \textbf{#1} (#2) #3}   
\newcommand{\HPA}[3]{\emph{ Helv.~Phys.~Acta} \textbf{#1} (#2) #3}

\newcommand{\JHEP}[3]{\emph{ JHEP} \textbf{#1} (#2) #3}  
\newcommand{\Tr}{\mathop{\rm Tr}}

\newcommand{\tR}{\tilde{t}_R}
\newcommand{\hS}{\widetilde{\mathcal{H}}}
\title{   
\vspace*{-1.3cm}   
\begin{flushright}   
\normalsize{
LPTENS-00/20\\      
IEM-FT-204/00\\
IFT-UAM/CSIC-00-31\\
FERMILAB-PUB-00/240-T \\
ANL-HEP-PR-00-101 \\
EFI-2000-030  \\
}
\end{flushright}    
\vspace{.5cm}
\Large
\textbf{Supersymmetric CP-violating Currents and\\
Electroweak Baryogenesis}
\vspace*{.2cm}
\author{\large\textbf
{M.~Carena$^a$, J.M.~Moreno$^b$, M.~Quir\'os$^{a,b}$}\\ \\
\textbf{
M.~Seco$^{b,\dagger}$} and \textbf{C.E.M.~Wagner$^{c,d}$}\\ \\
$^a$\normalsize\emph{Fermi National Accelerator Laboratory,
P.O. Box 500, Batavia, IL 60510, USA }\\
$^b$\normalsize\emph{Instituto de Estructura de la Materia (CSIC),
Serrano 123, 
E-28006 Madrid, Spain}\\
$^c$\normalsize\emph{HEP Division, Argonne National Laboratory,
9700 Cass Ave.,
Argonne, IL 60439, USA} \\
$^d$\normalsize\emph{Enrico Fermi Institute, Univ. of Chicago, 5640
Ellis Ave., Chicago, IL 60637, USA}}}
\date{}   
\begin{document}
\maketitle


\begin{abstract}
In this work we compute the CP-violating
currents of the right-handed stops and Higgsinos,
induced by the presence of non-trivial vacuum expectation
values of the Higgs fields within the context of the
minimal supersymmetric extension of the Standard Model (MSSM) 
with explicit CP-violating phases. Using the Keldysh formalism,
we perform  the computation of the currents
at finite temperature,
in an expansion of derivatives of the Higgs fields. 
Contrary to previous works, we implement a resummation
of the Higgs mass insertion effects to all orders in 
perturbation theory. While the components of the
right-handed stop current $j^\mu_{\widetilde t_R}$  
become proportional to the difference  
$H_2 \partial^{\mu}H_1-H_1 \partial^{\mu} H_2$
(suppressed by $\Delta\beta$), the Higgsino
currents, $j^\mu_{\widetilde{H}_i},$  
present contributions proportional to both 
$H_2 \partial^{\mu}H_1\pm H_1 \partial^{\mu} H_2$. 
For large values of the charged Higgs mass and
moderate values of $\tan\beta$ the contribution to the source
proportional to $H_2 \partial^{\mu}H_1+H_1 \partial^{\mu} H_2$
in the diffusion equations become sizeable,
although it is suppressed by  the Higgsino number
violating interaction rate $\Gamma_\mu^{-1/2}$.
For small values of the wall velocity,
$0.04\simlt v_\omega \simlt 0.1$, 
the total contribution leads to acceptable values
of the baryon asymmetry for values of the CP-violating phases $\varphi_{CP}$  
in the range $0.04\simlt|\sin\varphi_{CP}|\simlt 1$. Finally, we
comment on the relevance of the latest results of Higgs searches at LEP2
for the mechanism of electroweak baryogenesis within the MSSM. 
\end{abstract}
\vspace{.5cm}   

\hrule\vspace{1mm}
$^\dagger$\normalsize\emph{Present address: Department of Physics, 
University of Virginia, 382 McCormick Road, 
P.O. Box 400714, Charlottesville, VA 22904-4714.} 

\thispagestyle{empty}
\newpage

\section{Introduction} 
\label{introduction}
The origin of the baryon asymmetry of the Universe is one of the most
important open questions in cosmology and particle physics. It has
been long understood that, in order to generate the observed 
baryon asymmetry, three requirements~\cite{baryogenesis} 
need to be fulfilled: the 
non-conservation of baryon number, CP-violation and the existence
of non-equilibrium processes~\cite{reviews}. 
Interestingly enough, at temperatures
above the electroweak phase transition temperature, $T_c$, the Standard 
Model fulfills
these requirements. Baryon number violation is induced
by anomalous~\cite{anomaly} sphaleron processes~\cite{sphalerons}, 
which are suppressed
at zero temperature, but whose rate grows linearly with the
temperature above $T_c~\cite{sphalT}$. 
The non-conservation of CP is an essential property of the Standard 
Model, and non-equilibrium processes may be obtained through
the expansion of bubbles of true vacuum, which occurs after the 
electroweak phase transition.

In spite of fulfilling all the desired properties, the 
rate of the CP-violating processes in the Standard Model (SM)
is too small to induce the required baryon asymmetry~\cite{fs,huet}.
Moreover, the preservation of the generated baryon asymmetry
after the electroweak phase transition requires a strongly
first order phase 
transition~\footnote{An alternative dynamics for preserving the
generated baryon asymmetry has been explored in Ref.~\cite{marcelo}.}, 
with $v(T_c)/T_c \simgt 1$,
where $v(T_c)$ is the Higgs vacuum expectation value at
the critical temperature $T_c$. For the experimentally allowed
values of the Higgs mass, this requirement is not
fulfilled in the Standard Model~\cite{SMpt}. 

Supersymmetric particles lead to new radiative corrections
to the Higgs effective potential at finite 
temperature~\cite{early}-\cite{mariano2}. Light
boson fields with relevant couplings to the Higgs field may induce
a stronger first order electroweak phase 
transition~\cite{CQW}-\cite{LR}. The
supersymmetric partners of the top quark are the only new
bosons which couple in a relevant way to the Higgs boson
which acquire vacuum expectation value and hence play a relevant
role in defining the strength of the phase transition~\footnote{
Although bottom and tau Yukawa couplings become large  
for large values of $\tan\beta$, the bottom and tau superfield
couplings to the Higgs boson combination which acquires vacuum 
expectation value, $\Phi = H_1^0 \cos\beta + H_2^0 \sin\beta$, 
remains small, apart from an enhancement of the $\Phi$-trilinear coupling
to left and right sbottoms and staus, 
which increases the corresponding mixings, but does not
lead to an enhancement of the phase transition strength.}.
For sufficiently small values of the stop masses the strength
of the phase transition is enhanced~\cite{CQW,CQW2}. 
In order to get values
of $v(T_c)/T_c \geq 1$, however, the right handed stop 
soft supersymmetry breaking squared mass
parameter, $m_U^2$, should be small or even slightly negative and the
stop mixing mass parameter $X_t = |A_t - \mu_c/\tan\beta|$ must
be smaller than $\sim$ 0.6 $m_Q$, with $m_Q$ the left-handed stop supersymmetry
breaking mass parameter.  Under these conditions, and for 
$m_Q \simlt $1--3 TeV, a strongly 
first order phase transition may be obtained up to values of the 
lightest CP-even Higgs boson mass as high as 
$\sim$ 110--115 GeV~\cite{CQW2,LR}.

Moreover, supersymmetric particles lead to new, relevant
CP-violating sources for the generation 
of the baryon asymmetry~\cite{CPviol}.
Several computations have been performed~\cite{hn}-\cite{plus} 
in recent years, showing
that if the CP-violating phases associated with the chargino
mass parameters are not too small, these sources may lead to 
acceptable values of the baryon asymmetry. In this
work, we shall perform a computation of these
new CP-violating sources in an
expansion in derivatives of the Higgs background fields. 
Similarly to Ref.~\cite{CQRVW}, we shall use the Keldysh 
formalism~\cite{keldysh} for the computation of the CP-violating currents
at finite temperature. We improve the computation of 
Ref.~\cite{CQRVW} in two main aspects. On the one hand, 
instead of computing the temporal
component of the current in the lowest order of Higgs 
background insertions, we compute all current components by 
performing a resummation of the 
Higgs mass insertion contributions to all order in perturbation
theory. The resummation is essential since it leads to a
proper regularization of the resonant contribution to the
temporal component of the current found in Ref.~\cite{CQRVW}
and leads to contributions which are not suppressed for
large values of the charged Higgs mass. On the other hand, we consider, in
the diffusion equations, the contribution of Higgsino number violating
interaction rate~\cite{plus} from the Higgsino $\mu$ term in the lagrangian, 
$\Gamma_\mu$, that was considered in our previous calculations in the limit
$\Gamma_\mu/T\to\infty$. 

This article is organized as follows. In sections 
\ref{squark} and \ref{chargino} 
we present the detailed derivation of the CP-violating currents 
for the cases of right-handed top squarks ($j^\mu_{\tR}$) and charginos
($j^\mu_{\hS}$), respectively, 
by making use of the Keldysh formalism and resumming to all order 
in Higgs background insertions. These two sections deal with all the technical
details of the computation, with the main results
given in
Eqs.~(\ref{corrf}), (\ref{jota+}) and (\ref{jota-}).
In section \ref{bau} we present explicit, analytical, solutions to the
diffusion equations and an explicit expression for the baryon asymmetry in
the broken phase after the phase transition in the MSSM. 
In section~\ref{numerical} we exhibit the results of a numerical analysis of
our solutions. A discussion of present Higgs mass constraints is made
in section~\ref{higgs}, and in section \ref{conclusion} we present our 
conclusions and outlook.

\section{The squark sector}
\label{squark}

Our aim in this section is to compute the 
Green functions for left-handed 
($\widetilde t_L(x)$) and 
right-handed ($\widetilde t_R(x)$) stop fields,
describing the propagation of these scalars in the presence
of a bubble wall. The bubble wall is assumed to be located at the
space-time point $z$, where there is a non-trivial 
background of the MSSM Higgs fields, $H_i(z)$, which carries
dimensionful CP-violating couplings to the left- and
right-handed stops. We shall use these Green functions to
compute the right-handed and left-handed stop currents at the point $z$.
The starting point is the lagrangian for the stop system:
\begin{equation}
\label{lagstop}
{\cal L}(x)=\left|\partial_\mu \widetilde t_L(x)\right|^2+
\left|\partial_\mu \widetilde t_R(x)\right|^2+\left(
\begin{array}{ll}
\widetilde t^{\, *}_L(x) & \widetilde t^{\, *}_R(x)
\end{array}
\right)
{\cal M}(x)\left(
\begin{array}{l}\widetilde t_L(x)\\ 
\widetilde t_R(x)
\end{array}\right) \ ,
\end{equation}
where ${\cal M}$ is the stop squared mass matrix which depends, 
through the Higgs
background, on the space-time point.

Clearly this is not a free lagrangian, since the mass matrix depends on the 
space-time coordinates, and we must identify the free and perturbative parts
out of it. 
In order to make such a selection we will expand the mass matrix around the
point $z^\mu\equiv (\vec{r},t)$ (the point where we are calculating the 
currents in the plasma frame) up to first order 
in derivatives as,
\begin{equation}
\label{Mexp}
{\cal M}(x)={\cal M}(z)+(x-z)^\mu {\cal M}_\mu(z) \ ,
\end{equation}
where we use the notation ${\cal M}_\mu(z)\equiv\partial {\cal M}(z)/\partial z^\mu$, 
and we can split the initial Lagrangian as:
\begin{eqnarray}
\label{lagdec}
{\cal L}_0(x)&=&\left|\partial_\mu \widetilde t_L(x)\right|^2+
\left|\partial_\mu \widetilde t_R(x)\right|^2+\left(
\begin{array}{ll}
\widetilde t^{\, *}_L(x) & \widetilde t^{\, *}_R(x)
\end{array}
\right)
{\cal M}(z)\left(
\begin{array}{l}\widetilde t_L(x)\\ 
\widetilde t_R(x)
\end{array}\right)
\nonumber\\
{\cal L}_{\rm int}&=&(x-z)^\mu\left(
\begin{array}{ll}
\widetilde t^{\, *}_L(x) & \widetilde t^{\, *}_R(x)
\end{array}
\right)
{\cal M}_\mu (z)\left(
\begin{array}{l}\widetilde t_L(x)\\ 
\widetilde t_R(x)
\end{array}
\right)\ .
\end{eqnarray}

Let ${\cal U}(z)\in SU(2)$ be the matrix that diagonalizes 
${\cal M}(z)$. We can then rewrite ${\cal L}_0$ and ${\cal L}_{\rm int}$ as:
\begin{eqnarray}
\label{lagdec2}
{\cal L}_0&=&\sum_{i=1}^2\left\{\left|\partial_\mu \chi_i(x)\right|^2
+m_i^2(z)\left|\chi_i(x)\right|^2 \right\}\ ,\nonumber\\
{\cal L}_{\rm int}&=&(x-z)^\mu\left(
\begin{array}{ll}
\widetilde t^{\, *}_L(x) & \widetilde t^{\, *}_R(x)
\end{array}
\right)
{\cal U}(z){\cal M}_\mu (z)\ {\cal U}^\dagger(z)
\left(
\begin{array}{l}\widetilde t_L(x)\\ 
\widetilde t_R(x)
\end{array}
\right)\ ,
\end{eqnarray}
where $m^2_i(z)$, $\chi_i(z)$ $(i=1,2)$ are the eigenvalues 
and eigenvectors of $M(z)$. Note that the description in 
terms of the mass eigenstates $\chi_i(z)$ is useful so far
the Higgs field variations are small for propagation lengths
of the order of the inverse of the width of the stop fields, $\Gamma^{-1}$. 
Under  these conditions, namely $L_w \Gamma/v_w \simgt 1$, with
$L_w$ and $v_w$ being the bubble wall width and velocity, respectively, 
an expansion in derivatives is justified~\cite{CQRVW}. 

Now we can write down the two point Green function for the field 
$\left(\chi_1(x)\ \chi_2(x)\right)^T$ in matrix form:
\begin{equation}
\label{green}
G^{\chi}(x,y;z)=G(x,y;z)+\int{d^4w\ (w-z)^\mu G(x,w;z)\
{\cal U}(z)\, {\cal M}_\mu (z)\, {\cal U}^\dagger(z)\ G(w,y;z)
}+\ldots
\end{equation}
where  $x$ and $y$ are assumed to be close to $z$,
the point at which the current is being evaluated and around
which the expansion is being performed
($|x -z|,|y-z| \ll \Gamma^{-1}$), 
$G(x,y;z)$ is the two by two diagonal free Green function 
of the stop mass eigenstates with masses $m_i(z)$, the trace 
over internal ($a=1,2$) indices being understood in Eq.~(\ref{green}). 
Explicitly, the free Green functions for each of the two stop
eigenstates can be written as~\cite{keldysh}:
\begin{eqnarray}
\label{prop}
G_i^{11}&=&P_i^+ +f_B \left(P_i^+ -P_i^-\right)\nonumber\\
G_i^{12}&=&\left[\theta(p^0)+f_B\right] \left(P_i^+ -P_i^-\right)\nonumber\\
G_i^{21}&=&\left[\theta(-p^0)+f_B\right] \left(P_i^+ -P_i^-\right)\nonumber\\
G_i^{22}&=&-P_i^- +f_B \left(P_i^+ -P_i^-\right)\ ,
\end{eqnarray}
where $f_B\equiv n_B(|p^0|)$ is the Bose-Einstein distribution function, 
which contains the dependence on the temperature $T$, 
\begin{equation}
\label{pes}
P_i^{\pm}=\frac{1}{p_0^2-\vec{p}^2-m_i^2(z)\pm 2 i\Gamma_{\widetilde t}|p^0|}
\ ,
\end{equation}
and $\Gamma_{\widetilde t}$ is the stop width which can be taken to be 
$\Gamma_{\widetilde t}\,\sim \alpha_s\, T$ independently of the stop mass eigenstate. 

Since we need to calculate the CP-violating currents induced by
the right-handed stop states, we should first go to the weak
eigenstate basis. The Green functions in the weak eigenstate basis
can be obtained from the ones given above, which were computed in the
basis of mass eigenstates, by the following expression
$$G^{\widetilde t}(x,y;z)={\cal U}^\dagger(z)G^{\chi}(x,y;z){\cal U}(z)
\ .$$

Therefore, the current for right-handed stops takes 
the form:
\begin{equation}
\label{corriente}
j_{\tR}^\mu(z)=\lim_{x,y\to z}{\rm Tr}\left[P_2 \frac{\partial G^{\widetilde t}(x,y;z)}
{\partial (x-y)_\mu}\right] \ ,
\end{equation}
where $P_2 =(\sigma_0-\sigma_3)/2$, 
$\sigma_i$ being the two by two Pauli matrices and $\sigma_0$ 
the two by two identity matrix,
is a projection matrix which 
allows to separate the current induced by the right-handed stops
from the one induced by the left-handed stops. Nevertheless, 
since baryon number is conserved at this point, the total CP-violating
currents induced by left- and right-handed top squarks must be zero,
${\rm Tr}[\partial^\mu G^{\widetilde t}(x,y)]=0$, and therefore 
\begin{equation}
\label{current}
j_{\tR}^\mu(z)=-\frac12\lim_{x,y\to z}{\rm Tr}\left[\sigma_3
\frac{\partial G^{\widetilde t}(x,y;z)}
{\partial (x-y)_\mu}\right] \ .
\end{equation}
After integrating over the $w$ space-time variable, and going to momentum
space, we can write 
the current in terms of free Green functions of the mass eigenstates
at the point $z$:
\begin{eqnarray}
\label{corr2}
j_{\tR}^\mu(z)&=&\frac{1}{2}\int\frac{d^4p}{(2\pi)^4}p^\mu{\rm Tr}
\left[\sigma_3\,
{\cal U}^\dagger(z)\,G^{\nu}(p;z)\,
{\cal U}(z)\,{\cal M}_\nu(z)\,
{\cal U}^\dagger(z)\,G(p;z){\cal U}(z)\right.\nonumber\\
&-&\left.\sigma_3\,{\cal U}^\dagger(z)\,G(p;z)\,{\cal U}(z)\,
{\cal M}_\nu (z)\,
{\cal U}^\dagger(z)\, G^{\nu}(p;z)\,
{\cal U}(z)\right]
\end{eqnarray}
since the contribution induced by the linear term in $z$ in Eq. (\ref{green})
trivially vanishes because $G(p;z)$ only depends on 
$|{\bf p}|$ and $p^0$. We are using the notation $G^\nu(p;z)=\partial
G(p;z)/\partial p_\nu$.
Note that in the above expression only off-diagonal terms of the
derivatives of the mass matrix ${\cal M}_\nu(z)$ at the point $z$
give a non-vanishing contribution. We shall denote by 
$\widetilde{{\cal M}}_\nu(z)$,
the matrix containing only the derivative of the off-diagonal terms
of the matrix ${\cal M}(z)$.

The current could be simplified a little bit more by using:
\begin{equation}
\label{identity}
{\cal U}^{\dagger}(z)D\,{\cal U}(z)=\sigma_1D
\sigma_1+\frac12{\rm Tr}[{\cal U}(z)]
{\rm Tr}[D\sigma_3]{\cal U}^{\dagger}(z)\sigma_3
\end{equation}
where $D$ is a diagonal matrix. Then $j_{\widetilde t_R}^\mu(z)$ can be written as:
\begin{align}
\label{simcurr}
j_{\tR}^\mu(z)=&-\frac{i}{4}{\rm Tr}[{\cal U}(z)]{\rm Tr}
\left[\widetilde{{\cal M}}_\nu (z)
{\cal U}(z)\right]\int\frac{d^4p}{(2\pi)^4}p^\mu
{\rm Tr}\left[\sigma_1G(p;z)\sigma_2
G^{\nu}(p;z)\right]\nonumber\\
=&\frac{1}{4}{\rm Tr}[{\cal U}(z)]{\rm Tr}
\left[\widetilde{{\cal M}}_\nu (z)
{\cal U}(z)\right]\int\frac{d^4p}{(2\pi)^4}p^\mu\epsilon^{ij}
G_i(p;z)\ G^{\nu}_j(p;z)\ .
\end{align}
Expanding $G_i$ in terms of $P^\pm_i$ one gets:
\begin{align}
\label{currexp}
j_{\tR}^i(z)=&\frac{8\,C^i}{3\pi}{\rm Im}\left\{
\int_{-\infty}^\infty \frac{dp^0}{2\pi}
(1+2f_B)\int_0^{\infty}\frac{d{\bf p}}{2\pi}~{\bf p}^4
(P_1^+(p;z)P_2^+(p;z))^2\right\}
\nonumber\\
j_{\tR}^0(z)=&-\frac{2\,C^0}{\pi}
{\rm Im}\left\{\int_{-\infty}^\infty\frac{dp^0}{2\pi}|p^0|\left[(1+2f_B)
\left(|p^0|+i\Gamma_{\widetilde t}\right)\int_0^{\infty}\frac{d{\bf p}}{2\pi}
\left(P_1^+(p;z)P_2^+(p;z)\right)^2\right.
\right.\nonumber\\
-&\left.\left.\int_0^\infty\frac{d{\bf p}}{2\pi}\,  
\frac{f'_B}{m_1^2\,(z)-m_2^2\,(z)}
\,P_1^+(p;z)P_2^-(p;z)\vphantom{{P^+}^2}\right]
\vphantom{\frac12}\right\}
\end{align}
where $f'_B$ is the derivative of $f_B$ with respect to its argument
and $C_\mu$ is given by
\begin{equation}
\label{constant}
C_\mu=(m_1^2\,(z)-m_2^2\,(z))\,{\rm Tr}[{\cal U}(z)]{\rm Tr}
\left[\widetilde{{\cal M}}_\mu(z)\,
{\cal U}(z)\right]\ .
\end{equation}

Using now the particular value of the squared mass matrix ${\cal M}$ for the stop 
system, 
\begin{equation}
{\cal M}(z)=\left(
\begin{array}{cc}
m_Q^2+h_t^2\,H_2^2(z) & h_t\left( A_t H_2(z)-\mu^*_c H_1(z)\right) \\
h_t\left( A_t^* H_2(z)-\mu_c H_1(z)\right) & m_U^2+h_t^2\,H_2^2(z)
\end{array}
\right)\ ,
\end{equation}
where $h_t$ is the top-quark Yukawa coupling, $A_t$ the left-right stop mixing
parameter, and $\mu_c$
the complex Higgsino mass parameter,
defined as $\mu_c\equiv \mu\exp(i\varphi_\mu)$, 
with $\mu$ real (positive or negative). 
In the above, we have neglected 
corrections $\mathcal{O}(g^2)$. In this approximation, 
the above constant vector $C_\mu$, Eq.~(\ref{constant}), 
can be written as:
\begin{equation}
\label{Csquark}
C^\mu=2\, h_t^2\ {\rm Im}(A_t\,\mu_c)\,\left\{H_2(z) H^\mu_1(z)-
H_1(z) H^\mu_2(z)\right\}\, .
\end{equation}

Hence, in order to compute the CP-violating currents induced by the
stop fields, the momentum integrals should be performed. Due to the
form of the free Green functions, Eq.~(\ref{prop}), the 
integral over the temporal component of the momentum cannot be
performed by standard integration methods in the complex plane.
It is therefore better to perform the integration over the spatial
components of the momentum and express the results as an integral
function over $p^0$, which admits a simple physical interpretation.
In order to perform the integrals of the spatial components of the
momentum, one should note that all functions depend only on $|{\bf p}|^2$.
Therefore, the angular integration can be trivially performed and
the integral over the modulus $|p|$ of a function ${\cal F}(|p|)$ can 
be written as half 
the integral on the whole real plane of the function ${\cal F}(x)$,
with ${\cal F}(x) = {\cal F}(-x)$. Doing this, we can perform the
spatial momentum integrals
in Eq.~(\ref{currexp}) by means of standard
techniques of integration in the complex plane and the residues theorem,
and we can cast the resulting currents as:
\begin{equation}
\label{corrf}
j_{\tR}^\mu(z)=h_t^2\ {\rm Im}(A_t\,\mu_c)\,\left\{H_2(z)H^\mu_1(z)-
H_1(z)H^\mu_2(z)\right\}
\left\{ {\cal F}_B(z)+\delta^{\mu\,0}{\cal G}_B(z)\right\}
\end{equation}
where 
\begin{align}
\label{integrales}
{\cal F}_B(z)=&\frac{1}{6\pi^2}{\rm Re}\int_0^\infty dp^0\,(1+2\, f_B)
\left(\frac{1}{z_1+z_2}\right)^3
\nonumber\\
{\cal G}_B(z)=&\frac{1}{3\pi^2}{\rm Re}\int_0^\infty\,dp^0\,  p^0 f'_B
\left\{\left(\frac{1}{z_1+z_2}\right)^3\right.\nonumber\\
-&\left.\frac{3}{m_1^2\,(z)-m_2^2\,(z)}
\left[\frac{z_1}{m_1^2-m_2^2-4 i \Gamma_{\widetilde t}\ p^0}+
\frac{z_2}{m_1^2-m_2^2+4 i \Gamma_{\widetilde t}\ p^0}\right]\right\}
\end{align}
and $z_i$ is defined as the pole of $P_i^+$, i.e.
\begin{equation}
\label{zetas}
z_i(p^0)=\sqrt{p^0\left(p^0+2\, i \Gamma_{\widetilde t}\right)-m_i^2(z)}
\end{equation}
with positive real and imaginary parts satisfying ${\rm Re}(z_i)
=\Gamma_{\widetilde t}\ p^0/{\rm Im}(z_i)$. 

\section{The chargino sector}
\label{chargino}

For the case of the charged gaugino-Higgsino system
we will follow similar steps as the ones we performed before to
compute the stop current.
In this case the role of right-handed stops is played by the 
(left- and right-handed) Higgsinos. The starting point is the lagrangian:

\begin{equation}
\label{charlag}
\mathcal{L}(x)=\overline{\widetilde{h}}_c(x)\partial_\mu\gamma^\mu
\widetilde{h}{}_c(x)+
\overline{\widetilde{W}}_c(x)\partial_\mu\gamma^\mu\widetilde{W}_c(x)+
\begin{pmatrix}
\overline{\widetilde{W}}_c(x) & \overline{\widetilde{h}}_c(x)
\end{pmatrix}
M(x)
\begin{pmatrix}
\widetilde{W}_c(x) \\ 
\widetilde{h}_c(x)
\end{pmatrix}
\end{equation}
where
\begin{align}
\widetilde{h}_c=
\begin{pmatrix}
\widetilde{h}_2^+ \\
{\widetilde{h}{}_1^-}^*
\end{pmatrix},\qquad &
\widetilde{W}_c=
\begin{pmatrix}
\widetilde{W}^+ \\
{\widetilde{W}{}^-}^*
\end{pmatrix}\notag \ .
\end{align}
From the structure of the chargino mass matrix we can write the lagrangian in 
the following form:
\begin{align}
\label{psilag}
\mathcal{L}(x)=&\psi_R(x)^\dagger\sigma_\mu\partial^\mu\psi_R(x)+
\psi_L(x)^\dagger\overline{\sigma}_\mu\partial^\mu\psi_L(x)\nonumber\\
+&
\psi_R(x)^\dagger M(x)\psi_L(x)+\psi_L(x)^\dagger M^{\dagger}(x)\psi_R(x)
\end{align}
where in this expression we have used
\begin{align}
\psi_R(x)=
\begin{pmatrix}
\widetilde{W}^+\\\widetilde{h}_2^+
\end{pmatrix},
&\qquad
\psi_L(x)=
\begin{pmatrix}
\widetilde{W}^-\\\widetilde{h}_1^-
\end{pmatrix}\notag \ .
\end{align}

Expanding the masses around the point $z$ and splitting the 
lagrangian into a free and a 
perturbative part, we can write, to first order in derivatives:
\begin{align}
\label{splitlag}
\mathcal{L}_0(x)=&\psi_R(x)^{\dagger}\sigma_\mu\partial^\mu\psi_R(x)+
\psi_L(x)^{\dagger}\overline{\sigma}_\mu\partial^\mu\psi_L(x)\nonumber\\
+&
\psi_R(x)^{\dagger}M(z)\psi_L(x)+\psi_L(x)^{\dagger}M^{\dagger}(z)\psi_R(x)
\nonumber\\
&\nonumber\\
\mathcal{L}_{int}(x)=&(x-z)^\mu\left\{
\psi_R(x)^{\dagger}M_\mu(z)\psi_L(x)+
\psi_L(x)^{\dagger}M_\mu^{\dagger}(z)\psi_R(x)\right\} \ .
\end{align}

Like for the scalar case we will diagonalize $M(z)$ by means of the matrices 
$\mathcal{U}(z)$, $\mathcal{V}(z)\in SU(2)$. Additional phase
redefinition can be performed in order to bring the mass eigenstates
to be real and positive. In general, the 
lagrangian can be written as:
\begin{align}
\label{philag}
\mathcal{L}_0(x)=&\varphi_R(x)^{\dagger}\sigma_\mu\partial^\mu\varphi_R(x)+
\varphi_L(x)^{\dagger}\overline{\sigma}_\mu\partial^\mu\varphi_L(x)\notag\\
+&\varphi_R(x)^{\dagger}\begin{pmatrix}
m_1(z)&0\\
0&m_2(z)
\end{pmatrix}
\varphi_L(x)+\varphi_L(x)^{\dagger}\begin{pmatrix}
m^*_1(z)&0\\
0&m^*_2(z)
\end{pmatrix}
\varphi_R(x)\nonumber\\
&\nonumber\\
\mathcal{L}_{int}(x)=&(x-z)^\mu \left\{
\varphi_R(x)^{\dagger}\mathcal{U}(z) M_\mu(z)
\mathcal{V}^{\dagger}(z)
\varphi_L(x)+\varphi_L(x)^{\dagger}\mathcal{V}(z) 
M^{\dagger}_\mu(z)\mathcal{U}^{\dagger}(z)
\varphi_R(x) \right\}
\end{align}
where $m_i(z)$ are the eigenvalues of $M(z)$ and
\begin{align}
\varphi_R(x)=\mathcal{U}(z)\psi_R(x),&\qquad
\varphi_L(x)=\mathcal{V}(z)\psi_L(x)\notag
\end{align}
are the mass eigenstates at the point $z$.

At this point we can write the  Green functions describing the 
propagation of the right- and left-handed fermion 
$\varphi$ fields, $S_\varphi^{RR}$ and $S_\varphi^{LL}$,
respectively, as
\begin{align}
\label{cgreen}
S_\varphi^{RR}(x,y;z)=&S^{RR}(x,y;z)\notag\\
+&\int{d^4w(w-z)^\mu S^{RR}(x,w;z)\,\mathcal{U}(z)\, M_{\mu}(z)
\mathcal{V}^{\dagger}(z)\,S^{LR}(w,y;z)}\notag\\
+&\int{d^4w(w-z)^\mu S^{RL}(x,w;z)\mathcal{V}(z)\,
M^{\dagger}_{\mu}(z)\mathcal{U}^{\dagger}(z)S^{RR}(w,y;z)}
\notag\\
S_\varphi^{LL}(x,y;z)=&S^{LL}(x,y;z)\notag\\
+&\int{d^4w(w-z)^\mu S^{LL}(x,w;z)\mathcal{V}(z)\, 
M^{\dagger}_{\mu}(z)\mathcal{U}^{\dagger}(z)S^{RL}(w,y;z)}\notag\\
+&\int{d^4w(w-z)^\mu S^{LR}(x,w;z)\mathcal{U}(z)\, M_{\mu}(z)
\mathcal{V}^{\dagger}(z)\,S^{LL}(w,y;z)}
\end{align}
where $S^{LL}$, $S^{RR}$, $S^{LR}$ and $S^{RL}$ denote the left-left,
right-right, left-right and right-left Green functions of free 
fermions with mass $m_i(z)$.
In the approximation where both fermionic widths are equal,
we can rewrite the free
fermionic Green functions in terms of bosonic ones as:
\begin{xalignat}{2}
\label{relations}
S^{RR}(p;z)=&\sigma_\mu p^\mu G(p;z)&\qquad 
S^{RL}(p;z)=&\begin{pmatrix}m_1(z)&0\\0&m_2(z)\end{pmatrix} G(p;z)
\notag\\
S^{LR}(p;z)=&\begin{pmatrix}m^*_1(z)&0\\0&m^*_2(z)\end{pmatrix} G(p;z)
&\qquad S^{LL}(p;z)=&\overline{\sigma}_\mu p^\mu G(p;z)
\end{xalignat}
where the free Green functions $G(p;z)$ are given by (\ref{prop}) with 
$f_B\to f_F\equiv -n_F(|p^0|)$, $n_F$ being the Fermi-Dirac distribution 
function, $m_i(z)\to \left|m_i(z)\right|$, and
$\Gamma_{\widetilde{t}}\,\to\Gamma_{\widetilde{H}}\,\sim \alpha_W\, T$. 
Using the relations between  Green 
functions in the mass and weak eigenstate basis, as we did in the stop 
case, we obtain in the weak eigenstates basis,
\begin{xalignat}{2}
\label{psi-phi}
S_\psi^{RR}(p;z)=&\,\mathcal{U}^{\dagger}(z)S_\varphi^{RR}(p;z)\,
\mathcal{U}(z)&\qquad 
S_\psi^{RL}(p;z)=&\,\mathcal{U}^{\dagger}(z)S_\varphi^{RL}(p;z)\,\mathcal{V}(z)
\notag\\
S_\psi^{LR}(p;z)=&\,\mathcal{V}^{\dagger}(z)S_\varphi^{LR}(p;z)\,\mathcal{U}(z)
&\qquad S_\psi^{LL}(p;z)=&\,\mathcal{V}^{\dagger}(z)S_\varphi^{LL}(p;z)\,
\mathcal{V}(z)\ .
\end{xalignat}
The Higgsino currents can now be defined as:
\begin{equation}
\label{chcurr}
j_{\hS_{\pm}}^\mu(z)=\lim_{x,y\to z}\left\{\Tr\left[P_2\sigma^\mu 
S_\psi^{RR}(x,y;z)]
\right]\pm\Tr\left[P_2\overline{\sigma}^\mu S_\psi^{LL}(x,y;z)]\right]\right\}
\end{equation}
where $P_2$ is the projection operator used in Eq.~(\ref{corriente}).

By replacing (\ref{relations}) and (\ref{psi-phi}) in these currents, 
and taking into account that, as happened in the stop case,  
the contribution of the linear term in $z$ in Eq.~(\ref{cgreen})
is zero by symmetry reasons, one gets~\footnote{Notice that the phases
$\varphi_i$ of the mass eigenvalues, $m_i(z)=|m_i(z)|\exp\{i\varphi_i(z)\}$
can be absorbed in a redefinition of the matrix
$\mathcal{V}(z)$, as $\mathcal{V}(z)\to {\rm diag}(\exp\{i\varphi_1(z)\},
\exp\{i\varphi_2(z)\})\mathcal{V}(z)$. As required, the currents
(\ref{cor}) do not depend on this phase redefinition.} 
\begin{align}
j_{\hS_{\pm}}^\mu(z)=&\frac12
\int\frac{d^4p}{(2\pi)^4}\left\{p^\mu\Tr
\left[\sigma_3\left(\,\mathcal{U}^{\dagger}(z)
G(p;z)\,\mathcal{U}(z) M_\rho(z)M^{\dagger}(z)
\mathcal{U}^{\dagger}(z)G^\rho(p;z)\,
\mathcal{U}(z)\right.\right.\right.\notag\\
-&\,\mathcal{U}^{\dagger}(z)G^\rho(p;z)\,\mathcal{U}(z)M(z) 
M_\rho^{\dagger}(z)\,\mathcal{U}^{\dagger}(z)G(p;z)\,\mathcal{U}(z)\notag\\
\pm &\mathcal{V}^{\dagger}(z)G(p;z)\mathcal{V}(z) M_\rho^{\dagger}(z)M(z)
\mathcal{V}^{\dagger}(z)G^\rho(p;z)\mathcal{V}(z))\notag\\
\mp &\left.\left.\mathcal{V}^{\dagger}(z)G^\rho(p;z)\mathcal{V}(z)
M^{\dagger}(z) M_\rho(z)\mathcal{V}^{\dagger}(z)G(p;z)
\mathcal{V}(z)\right)\right]\nonumber\\
+&\Tr\left[\,\mathcal{U}^{\dagger}(z)G(p;z)\,\mathcal{U}(z)(
M^\mu(z)M^{\dagger}(z)-M(z) M^{\mu\,\dagger}(z))\,\mathcal{U}^{\dagger}(z)
G(p;z)\,\mathcal{U}(z)\right.\nonumber\\
\pm&\left.\left.\mathcal{V}^{\dagger}(z)G(p;z)\mathcal{V}(z)
(M^{\mu\,\dagger}(z)M(z)-M^{\dagger}(z) M^\mu(z))
\mathcal{V}^{\dagger}(z)G(p;z)\mathcal{V}(z)\right]\right\}\ .
\label{cor}
\end{align}

The chargino mass matrix is given by
\begin{equation}
\label{masach}
M(z)=\left(
\begin{array}{cc}
M_2 & u_2(z) \\
u_1(z) & \mu_c
\end{array}
\right)
\end{equation}
where we have defined $u_i(z)\equiv g H_i(z)$. 
The diagonalizing matrices are~\cite{CJK}
\begin{align}
\label{UV}
\mathcal{U}=&\frac{1}{\sqrt{2\, \Lambda(\Delta+\Lambda)}}
\left(
\begin{array}{cc}
\Delta+\Lambda & M_2\, u_1+\mu^*_c\, u_2 \\
-\left( M_2\, u_1+\mu_c\, u_2 \right) & \Delta+\Lambda
\end{array}
\right)
\nonumber\\
\mathcal{V}=&\frac{1}{\sqrt{2\, \Lambda(\bar\Delta+\Lambda)}}
\left(
\begin{array}{cc}
\bar\Delta+\Lambda & M_2\, u_2+\mu_c\, u_1 \\
-\left( M_2\, u_2+\mu^*_c\, u_1 \right) & \bar\Delta+\Lambda
\end{array}
\right)\ ,
\end{align}
where field redefinitions have been made in order to make
the Higgs vacuum expectation values, as well as the weak gaugino
mass $M_2$, real, 
\begin{align}
\label{defin}
\Delta=&(M_2^2-|\mu_c|^2-u_1^2+u_2^2)/2 \nonumber\\
\bar\Delta=&(M_2^2-|\mu_c|^2-u_2^2+u_1^2)/2 \nonumber\\
\Lambda=&\left(\Delta^2+\left|M_2\,u_1+\mu^*_c\, u_2 \right|^2\right)^{1/2}\ ,
\end{align}
and the mass eigenvalues are given by
\begin{align}
\label{eigenval}
m_1(z)=& \frac{\left(\Delta+\Lambda+u_1^2(z)\right)M_2+u_1(z) u_2(z)\mu^*_c}
{\sqrt{(\Delta+\Lambda)(\bar\Delta+\Lambda)}}
\nonumber\\
m_2(z)=& \frac{\left(\Delta+\Lambda-u_2^2(z)\right)\mu_c-u_1(z) u_2(z)M_2}
{\sqrt{(\Delta+\Lambda)(\bar\Delta+\Lambda)}} \ .
\end{align}

Using these expressions, and the property $2\,\Lambda=|m_1(z)|^2-|m_2(z)|^2$, 
we can cast the Higgsino currents in the following general form:
\begin{align}
\label{j+}
j^\mu_{\hS_{+}}=& \frac{{\rm Im}(M_2\, \mu_c)}{\Lambda}
\left\{\left[ u_2(z) u_1^\nu(z)-
u_1(z) u_2^\nu(z)\right]\int\frac{d^4 p}{(2\pi)^4}\,p^\mu\, G_i(p;z)
\epsilon^{ij}G_j^\nu(p;z) \right.\\
+& \left.\frac{u_2^2(z)-u_1^2(z)}
{2\,\Lambda}\left[ u_2(z) u_1^\nu(z)+
u_1(z) u_2^\nu(z)\right]\int\frac{d^4 p}{(2\pi)^4}\,p^\mu\, G_i(p;z)
\left(\delta^{ij}-\sigma_1^{ij}\right)G_j^\nu(p;z)\right\}\nonumber
\end{align}
where $\epsilon^{12}=+1$, and 
\begin{align}
\label{j-}
j^\mu_{\hS_{-}}=& \frac{{\rm Im}(M_2\, \mu_c)}{2\,\Lambda}
\left[ u_2(z) u_1^\nu(z)+u_1(z) u_2^\nu(z)\right]
\int\frac{d^4 p}{(2\pi)^4}\,p^\mu\,\nonumber\\
\times &\left\{ \left[(G_2+G_1)(G_2-G_1)\right]^\nu +\left(
\frac{\Delta+\bar\Delta}{\Lambda}\right)\left(G_2-G_1\right)
\left(G_2^\nu-G_1^\nu\right)\right\}\ .
\end{align}

Notice that while the first term in $j_{\hS_{+}}^\mu$ 
is similar to the squark current 
$j_{\tR}^\mu$ (it is proportional to  $u_2(z) u_1^\nu(z)-u_1(z) u_2^\nu(z)$), 
the second term in $j_{\hS_{+}}^\mu$ and the current $j_{\hS_{-}}^\mu$  
have no counterpart in the scalar sector. The contribution 
proportional to $u_2(z) u_1^\nu(z)-u_1(z) u_2^\nu(z)$
is proportional to the variation $\Delta\beta$
of the angle $\beta$ along the bubble wall. Since $\Delta\beta\simlt 10^{-2}$,
the corresponding contribution is suppressed. The contribution
proportional to $u_2(z) u_1^\nu(z)+u_1(z) u_2^\nu(z)$, 
instead, is not affected by this suppression factor, although it is suppressed,
for large values of $\tan\beta$, as $1/\tan\beta$.

Now the integration over the spatial components of the momentum
can be performed as in the previous section and the final
currents can be cast as follows. For the current $j_{\hS_{+}}^\mu$ one obtains,
\begin{align}
\label{jota+}
j_{\hS_{+}}^\mu(z)=& 2\,{\rm Im}(M_2\, \mu_c)\left\{
\left[u_2(z) u_1^\mu(z)-u_1(z) u_2^\mu(z)\right]\left\{\mathcal{F}_F(z)
+\delta^{\mu\,0}\mathcal{G}_F(z)\right\}\right.\nonumber\\
+&\left.\left[u_2^2(z)-u_1^2(z)\right]\left[u_2(z) u_1^\mu(z)+
u_1(z) u_2^\mu(z)\right]\mathcal{H}_F(z)\right\}
\end{align}
where the functions $\mathcal{F}_F,\ \mathcal{G}_F$ are defined in 
(\ref{integrales}) after changing $f_B\to f_F$, 
$m_i(z)\to \left|m_i(z)\right|$ and $\Gamma_{\widetilde{t}}
\to\Gamma_{\widetilde{H}}$, and
\begin{equation}
\label{hF}
\mathcal{H}_F(z)=\frac{1}{8\pi^2}{\rm Re}\int_0^\infty dp^0\,(1+2\, f_F)
\frac{1}{z_1\, z_2}\left(\frac{1}{z_1+z_2}\right)^3
\end{equation}
with $z_i(z)$  defined in (\ref{zetas}), after changing $m_i(z)\to
\left|m_i(z)\right|$ and $\Gamma_{\widetilde{t}}
\to\Gamma_{\widetilde{H}}$. 
Note that, being proportional to 
$u_2^2(z)-u_1^2(z)\equiv -u^2(z)\cos 2\beta(z)$, the second term of
$j^{\mu}_{\hS_{+}}$ vanishes at the lowest order in the Higgs field
insertions, in agreement with our previous results~\cite{CQRVW}, 
and it also vanishes in the case $\tan\beta=1$.

For the current $j_{\hS_{-}}^\mu$ one obtains,
\begin{equation}
\label{jota-}
j_{\hS_{-}}^\mu(z)=\,2\, {\rm Im}(M_2\, \mu_c)
\left[u_2(z) u_1^\mu(z)+u_1(z) u_2^\mu(z)\right]
\left\{\mathcal{K}_F(z)
+2\,\left[\Delta+\bar\Delta\right]\, \mathcal{H}_F(z)
\right\}
\end{equation}
where the function $\mathcal{K}_F$ is defined as,
\begin{equation}
\label{hK}
\mathcal{K}_F(z)=-\,\frac{1}{4\,\pi^2}{\rm Re}\int_0^\infty dp^0\,(1+2\, f_F)
\frac{1}{z_1\,z_2}\left(\frac{1}{z_1+z_2}\right)\ .
\end{equation}
Note that the current (\ref{jota-}) appears to leading order in the Higgs 
mass insertion and it is not suppressed by $\Delta\beta$. However, 
$j_{\hS_{-}}^\mu$ is suppressed for large values of $\tan\beta$ (which
are needed to push the Higgs mass beyond the most recent LEP bounds, as we
will discuss in section~\ref{higgs}) and, moreover, its effects on
the corresponding Higgs density are damped by the presence of
the Higgsino number violating interaction rate $\Gamma_\mu$. Accordingly,
its contribution to the BAU is small.

A similar calculation for the neutral gaugino-Higgsino system would involve
diagonalization of the four-by-four neutralino mass matrix, making the
analytic resummation treatment much more involved than for the chargino case.
The analysis performed in Ref.~\cite{CQRVW}, to lowest order in the Higgs
mass insertions, showed, as expected from a naive counting of degrees of
freedom, that the neutralinos contribute to the Higgsino current as
half the chargino contribution, with a total effect given by 3/2 
that of the chargino. After resummation of the neutralino sector it would be 
reasonable to expect a total contribution to the Higgsino current equal to
$\sim$ 3/2 that of the chargino sector. However, and to be as conservative as
possible in our calculation of the baryon asymmetry, we would just consider as
source of baryon number the chargino current (remember that left-handed
squarks are assumed very heavy and decouple from the thermal bath) that was
computed in this section, keeping in mind that an enhancement factor 
$\sim$ 3/2 might appear after a rigorous calculation of the currents in the 
neutralino sector. 

\section{The baryon asymmetry}
\label{bau}

To evaluate the baryon asymmetry generated in the broken phase we need
to first compute the density of left-handed quarks and leptons, $n_L$,
in front of the
bubble wall (in the symmetric phase). These chiral densities are
the ones that induce
weak sphalerons to produce a net
baryon number. Since, in the present scenario, there is essentially no lepton
asymmetry, the density to be computed in the symmetric 
phase~\footnote{We use, for the third
family, the notation $Q\equiv Q_3,\ T\equiv T_3$.} is 
$n_L=n_{Q}+\sum_{i=1}^2 n_{Q_i}$
where the density of a chiral supermultiplet 
$Q \equiv (q,\tilde q)$ 
is understood as the sum of densities of particle
components, assuming the supergauge interactions to be in thermal 
equilibrium, $n_Q=n_q+n_{\tilde q}$. If the system is near thermal equilibrium,
particle densities, $n_i$, are related to the local chemical potential,
$\mu_i$ by the relation $n_i=k_i \mu_i T^2/6$, where $k_i$ are statistical 
factors equal to 2 (1) for bosons (fermions) and exponentially suppressed
for particle masses $m_i$ much larger than $T$. For the calculation of
the density $n_L$ we will use the formalism described in 
Refs.~\cite{hn,CQRVW}.

We will consider those particle species that participate in fast particle 
number changing transitions, neglecting all Yukawa couplings except those
corresponding to the top quark. In this approximation, there is 
no left-handed lepton
number contribution to $n_L$. By introducing strong sphaleron effects, first
and second family quark number is generated. Assuming 
that all quarks have nearly
the same diffusion constant it turns out that~\cite{hn},
$n_{Q_1}=n_{Q_2}=2(n_{Q}+n_{T})$, and then,
\begin{equation}
\label{dobletes}
n_L=5\, n_Q+\, 4\, n_T \ .
\end{equation}

In general we will relate particle number changing, or fermion number
violating, rates $\Gamma_X$ with the corresponding rates per unit volume
$\gamma_X$, as,
\begin{equation}
\label{rates}
\Gamma_X=\frac{6\, \gamma_X}{T^3}\ .
\end{equation}
The involved weak and strong sphaleron rates are:
\begin{equation}
\Gamma_{ws}=\ 6\,\kappa_{ws}\, \alpha_w^5 T, \quad
\Gamma_{ss}=\ 6\,\kappa_{ss}\,\frac{8}{3}\, \alpha_s^4 T \ ,
\end{equation}
respectively,
where $\kappa_{ws}= 20\pm 2$~\cite{mr} and $\kappa_{ss}=\mathcal{O}(1)$.
The particle number changing rates that will be considered 
both in the symmetric and in the broken phase are: 
$\Gamma_{Y_2}$, corresponding to all supersymmetric and soft breaking
trilinear interactions arising from the $h_t H_2 Q T$ term in the 
superpotential, $\Gamma_{Y_1}$, which corresponds to the supersymmetric
trilinear scalar interaction in the Lagrangian involving the third generation
squarks and the Higgs $H_1$, and $\Gamma_\mu$, which corresponds to the 
$\mu_c \tilde{H}_1 \tilde{H_2}$ term in the Lagrangian. 
There are also the Higgs number
violating and axial top number violation processes, induced by the 
Higgs self interactions and by top quark mass effects,
with rates $\Gamma_h$
and $\Gamma_m$, respectively, that are only active in the broken phase.

We will write now a set of diffusion equations involving $n_Q$, $n_T$, 
$n_{H_1}$ (the density of $H_1\equiv(h_1,\tilde h_1)$) and $n_{H_2}$ 
(the density of $\bar H_2\equiv(\bar h_2,\tilde{\bar h}_2)$), and
the particle number changing rates and CP-violating source terms
discussed above.
In the bubble wall frame, and ignoring the curvature of the bubble wall,
all quantities become functions of $z\equiv r+ v_\omega t$, where $v_\omega$
is the bubble wall velocity. The diffusion equations are:
\begin{align}
\label{nQ}
v_\omega n'_Q= & D_q n''_Q-
\Gamma_Y\left[\frac{n_Q}{k_Q}-\frac{n_T}{k_T}-
\frac{n_H+\rho\, n_h}{k_H} \right]
-\Gamma_m\left[ \frac{n_Q}{k_Q}-\frac{n_T}{k_T} \right]\nonumber\\
-& 6 \Gamma_{ss} \left[2\,\frac{n_Q}{k_Q}-\frac{n_T}{k_T}+
9\,\frac{n_Q+n_T}{k_B} \right]+\tilde{\gamma}_Q
\end{align}
\begin{align}
\label{nT}
v_\omega n'_T= & D_q n''_T+
\Gamma_Y\left[\frac{n_Q}{k_Q}-\frac{n_T}{k_T}-
\frac{n_H+\rho\, n_h}{k_H} \right]
+\Gamma_m\left[ \frac{n_Q}{k_Q}-\frac{n_T}{k_T} \right]\nonumber\\
+& 3 \Gamma_{ss} \left[2\,\frac{n_Q}{k_Q}-\frac{n_T}{k_T}+
9\,\frac{n_Q+n_T}{k_B} \right]-\tilde{\gamma}_Q
\end{align}
\begin{align}
\label{nH}
v_\omega n'_H= & D_h n''_H+
\Gamma_Y\left[\frac{n_Q}{k_Q}-\frac{n_T}{k_T}-
\frac{n_H+\rho\, n_h}{k_H} \right]
-\Gamma_h\,\frac{n_H}{k_H}+\tilde{\gamma}_{\widetilde H_+}\\
\label{nh}
v_\omega n'_h= & D_h n''_h+
\rho\,\Gamma_Y\left[\frac{n_Q}{k_Q}-\frac{n_T}{k_T}-
\frac{n_H+\, n_h/\rho}{k_H} \right]
-\left(\Gamma_h+4\, \Gamma_\mu\right)\,
\frac{n_h}{k_H}+\tilde{\gamma}_{\widetilde H_-}
\end{align}
where all derivatives are with respect to $z$, 
$D_q\sim 6/T$ and $D_h\sim 110/T$ are the corresponding diffusion constants
in the quark and Higgs sectors~\cite{turok},
$n_H\equiv n_{H_2}+n_{H_1}$, $n_h\equiv n_{H_2}-n_{H_1}$,
$k_H\equiv k_{H_1}+k_{H_2}$, $\Gamma_Y\equiv \Gamma_{Y_2}+\Gamma_{Y_1}$
and $\rho\,\Gamma_Y\equiv \Gamma_{Y_2}-\Gamma_{Y_1}$. The parameter $\rho$ is
in the range $0\le\rho\le 1$. In previous analyses~\cite{hn,CQRVW,nuria} 
the limit $\Gamma_\mu\to\infty$ was implicitly considered, leading to the 
solution $n_h\to 0$. However, as we will see, for finite values of
$\Gamma_\mu$ we obtain non-vanishing values of the density $n_h$.

For the sources $\tilde{\gamma}_{Q,\hS_{\pm}}$ in Eqs.~(\ref{nQ})-(\ref{nh})
we will follow the formalism of Refs.~\cite{hn,Toni2} where $\tilde{\gamma}_X
\simeq j_X^0/\tau_X$, $\tau_X$ being the corresponding typical 
thermalization time. Thus we will use as sources of our diffusion equations,
\begin{align}
\label{sources}
\tilde{\gamma}_Q\simeq &-\, v_\omega\, h_t^2\, \Gamma_{\widetilde{t}}\ 
{\rm Im}(A_t\mu_c)\ H^2(z)\, \beta'(z)\,
\left\{ \mathcal{F}_B(z)+\mathcal{G}_B(z)
\right\}\nonumber\\
\tilde{\gamma}_{\widetilde H_+ }\simeq &-\, 2\, v_\omega\, g^2\, 
\Gamma_{\hS} \, {\rm Im}(M_2\mu_c)\left\{ H^2(z)\, \beta'(z)\,
\left[ \mathcal{F}_F(z)+\mathcal{G}_F(z)
\right]\right.\nonumber\\
+&\ \left. g^2\, H^2(z)\cos 2\beta(z)\left[
H(z) H'(z)\sin 2\beta(z)+H^2(z)\cos 2\beta(z) \beta'(z)
\right] \mathcal{H}_F(z) \right\}\nonumber\\
\tilde{\gamma}_{\widetilde H_-}\simeq &\ 2\, v_\omega\, g^2\, 
\Gamma_{\hS} \, {\rm Im}(M_2\mu_c)\left[
H(z) H'(z)\sin 2\beta(z)+H^2(z)\cos 2\beta(z) \beta'(z)
\right]\nonumber\\
&\left\{ \mathcal{K}_F(z)
+2\left(\Delta+\bar\Delta\right)\mathcal{H}_F(z)\right\} \ .
\end{align}
Notice that our sources, Eq.~(\ref{sources}), are proportional to the wall
velocity $v_\omega$, and so die when the latter goes to zero, which is
a physical requirement. 

We can find an approximate solution for $n_Q$ and $n_T$ by assuming that
$\Gamma_Y$ and $\Gamma_{ss}$ are fast so that 
$n_Q/k_Q-n_T/k_T-(n_H+\rho\, n_h)/k_H=\mathcal{O}
(1/\Gamma_Y)$ and $2\,n_Q/k_Q-n_T/k_T+
9\,(n_Q+n_T)/k_B=\mathcal{O} (1/\Gamma_{ss})$. In this case we can write
\begin{align}
\label{QT}
n_Q=&\ \frac{k_Q\left(9 k_T-k_B\right)}{k_H\left(k_B+9 k_Q
+9 k_T\right)}\
(n_H+\, \rho\, n_h)+\mathcal{O}\left(\frac{1}{\Gamma_{ss}}\, ,
\frac{1}{\Gamma_Y}\right)\nonumber\\ 
n_T=&-\ \frac{k_T\left(9 k_Q+2 k_B\right)}{k_H\left(k_B+9 k_Q
+9 k_T\right)}\
(n_H+\, \rho\, n_h)+\mathcal{O}\left(\frac{1}{\Gamma_{ss}}\, ,
\frac{1}{\Gamma_Y}\right)\ .
\end{align}
If the left-handed third generation squarks were light
($m_Q\sim T$) we could expect that 
all supersymmetric and supersymmetry breaking interactions arising 
from the $h_t\, H_2\, Q\, T$ term in
the superpotential are in thermal equilibrium and similar in size, so that
$\Gamma_{Y_1}\simeq \Gamma_{Y_2}$, or $\rho\ll 1$. In such case, which
was considered in Ref.~\cite{plus},  the influence
of $n_h$ in the quark densities $n_Q$ and $n_T$, 
through Eqs.~(\ref{QT}), is $\rho$-suppressed although 
this suppression can be arguably mild 
depending on the particularly chosen value of $\rho$.
However, in the case where left-handed squarks are heavy
($m_Q\gg T$), as preferred to get a good agreement of the MSSM 
with electroweak precision measurements, their corresponding 
interactions decouple, 
$\Gamma_{Y_1}\simeq0$ and $\rho\simeq 1$. 
This is the case we will consider from here on.

We now take (for $\rho=1$) 
the linear combinations of Eqs.~(\ref{nQ}), (\ref{nT}),  
(\ref{nH}) and (\ref{nh}) which are independent of $\Gamma_Y$ and 
$\Gamma_{ss}$. They are given by,
\begin{align}
\label{nH2}
v_\omega\left[n'_Q+2\, n'_T-n'_H\right]=& D_q\left[n''_Q+2\,n''_T\right]
-D_h\, n''_H
+\Gamma_m\left[\frac{n_Q}{k_Q}-\frac{n_T}{k_T} \right]\nonumber\\
+&\, \Gamma_h
\frac{n_H}{k_H}-\left(\tilde{\gamma}_Q+\tilde{\gamma}_{\widetilde{H}_{+}}
\right)\\
\label{nh2}
v_\omega\left[n'_Q+2\, n'_T-n'_h\right]=& D_q\left[n''_Q+2\,n''_T\right]
-D_h\, n''_h
+\Gamma_m\left[\frac{n_Q}{k_Q}-\frac{n_T}{k_T} \right]\nonumber\\
+&\,
\left[\Gamma_h+4\, \Gamma_\mu\right] 
\frac{n_h}{k_H}-\left(\tilde{\gamma}_Q+\tilde{\gamma}_{\widetilde{H}_{-}}
\right)\ .
\end{align}
When $n_Q$ and $n_T$ are replaced by the explicit solutions of Eqs.~(\ref{QT}),
as functions of $n_H$ and $n_h$, Eqs.~(\ref{nH2}) and (\ref{nh2}) yield 
the system of coupled equations for $n_H$ and $n_h$:
\begin{equation}
\label{sistema}
v_\omega\, \mathcal{A}
\left(
\begin{array}{c}
n'_H \\
n'_h
\end{array}
\right)= \mathcal{D}
\left(
\begin{array}{c}
n''_H \\
n''_h
\end{array}
\right)-\mathcal{G}
\left(
\begin{array}{c}
n_H \\
n_h
\end{array}
\right)+\left(
\begin{array}{c}
 f_{+}\\
f_{-}
\end{array}
\right)
\end{equation}
where the sources are
\begin{equation}
 f_{\pm}=\frac{G}{F+G}\,\left(
\tilde{\gamma}_Q+\tilde{\gamma}_{\widetilde{H}_{\pm}}\right)\ , 
\label{fuentes}
\end{equation}
with
\begin{align}
F\equiv&\ 9 k_Q k_T+k_Q k_B+ 4 k_T k_B\nonumber\\
G\equiv&\ k_H(9 k_Q+9 k_T+k_B)\ ,
\label{FG}
\end{align}
and $\mathcal{A}$, $\mathcal{D}$ and $\mathcal{G}$ are the $2\times 2$ 
matrices,
\begin{align}
\mathcal{A}=&\left(
\begin{array}{cc}
1 & \frac{F}{F+G}\\
\frac{F}{F+G} & 1
\end{array}\right) 
\ ,\ \mathcal{D}=\left(
\begin{array}{cc}
\overline{D}_q+\overline{D}_h &\overline{D}_q \\
\overline{D}_q & \overline{D}_q+\overline{D}_h
\end{array}\right) \nonumber\\ & \nonumber\\
\mathcal{G}=&\left(
\begin{array}{cc}
\overline{\Gamma}_m+\overline{\Gamma}_h &\overline{\Gamma}_h \\
\overline{\Gamma}_m & \overline{\Gamma}_m+\overline{\Gamma}_h
+4\, \overline{\Gamma}_\mu
\end{array}\right) \ ,
\label{matrices}
\end{align}
with
\begin{align}
\overline{D}_q\equiv & \frac{F}{F+G}\ D_q, \quad 
\overline{D}_h\equiv \frac{G}{F+G}\ D_h\nonumber\\ & \nonumber\\
\overline{\Gamma}_i\equiv & \frac{G}{F+G}\ \frac{\Gamma_i}{k_H},\quad
(i=m,\ h,\ \mu)\ .
\label{barras}
\end{align}

The system (\ref{sistema}) amounts to equations for $n_H$ and $n_h$, 
with  sources induced by
$\tilde{\gamma}_{\widetilde{Q}}$ and 
$\tilde{\gamma}_{\widetilde{H}_{\pm}}$, and by the same 
densities $n_{H,h}$ and
their derivatives. It can be re-written as,
\begin{align}
\label{nHfin}
v_\omega\, n'_H=& \overline D\, n''_H-\overline \Gamma\, n_H+f_{+}+\Delta f_+
\\
\label{nh3}
v_\omega\, n'_h=& D_h\, n''_h-\left[\Gamma_h+4 \Gamma_\mu\right]
\frac{n_h }{k_H}
+v_\omega\, n'_H- D_h\, n''_H+\Gamma_h\frac{n_H}{k_H}
+\tilde{\gamma}_{\widetilde{H}_{-}}
-\tilde{\gamma}_{\widetilde{H}_{+}}
\end{align}
where
\begin{align}
\label{Dbar}
\overline D=& \overline D_q+\overline D_h,\quad
\overline\Gamma= \overline\Gamma_m+\overline\Gamma_h\\
\label{deltaf}
\Delta f_+=& -\frac{F}{F+G}\, v_\omega n'_h+\overline D_q\, n''_h
-\overline\Gamma_m\, n_h \ .
\end{align}
We have solved the system (\ref{sistema}) numerically and the results are
presented in section~\ref{numerical}. However a very useful analytical
approximation can be worked out as follows. Using Eq.~(\ref{nH}) and the
approximate relations (\ref{QT}) we can
write for $n_h$ the following equation,
\begin{equation}
\label{nhfin}
v_\omega\, n'_h= D_h\, n''_h-\left[\Gamma_h+4 \Gamma_\mu\right]
\frac{n_h }{k_H}+\tilde{\gamma}_{\widetilde{H}_{-}}\ .
\end{equation}
In this way the equation for $n_h$ has been decoupled
from the other equations and can be easily solved. On the other hand, from
(\ref{nHfin}) and the expression for $\Delta f_+$, Eq.~(\ref{deltaf}), we see
that $n_h$ acts as a source for $n_H$, and the equations (\ref{nHfin})
and (\ref{nhfin}) can be solved analytically. 

We will only quote the solutions in the symmetric
phase ($z<0$) since that would be needed to compute the baryon asymmetry from
$n_L(z)$, as we will see. Finally we will impose boundary conditions
$n_h(\pm\infty)$ and $n_H(\pm\infty)$ and continuity of the functions and
first derivatives at $z=0$.

From Eq.~(\ref{nhfin}) we obtain the solution for $n_h(z)$, for $z\leq 0$ as,
\begin{equation}
\label{nhsol}
n_h(z)=\, \mathcal{A}_h\ e^{z\alpha_+}
\end{equation}
where
\begin{align}
\label{Ah}
\mathcal{A}_h=& \frac{2}{\sqrt{v_\omega^2+4\Gamma_1 D_h}
+\sqrt{v_\omega^2+4\Gamma_2 D_h}}
\int_0^\infty d\zeta\ \tilde{\gamma}_{\widetilde{H}_{-}}(\zeta)\,e^{-\zeta \beta_+}
\end{align}
and
\begin{align}
\label{alpha}
\alpha_{\pm}=& \frac{1}{2 D_h}\left\{v_\omega\pm
\sqrt{v_\omega^2+ 4\Gamma_1 D_h} \right\}\nonumber \\
\beta_\pm=& \frac{1}{2 D_h}\left\{v_\omega\pm
\sqrt{v_\omega^2+ 4\Gamma_2 D_h} \right\}\nonumber \\
\Gamma_2=&\frac{ \Gamma_h+4\Gamma_\mu}{k_H}\nonumber \\ 
\Gamma_1=& \frac{4\Gamma_\mu}{k_H}\ .
\end{align}
Note that, from expression (\ref{alpha}), the coefficient $\mathcal{A}_h$ 
behaves as $\Gamma_\mu^{-1/2}$, in the limit of large $\Gamma_\mu$, 
and so the $n_h$ density tends to zero when
$\Gamma_\mu$ tends to infinity, as anticipated.

From Eq.~(\ref{nHfin}), the solution for $n_H(z)$, for $z\leq 0$ is given by
\begin{equation}
\label{nHsol}
n_H(z)=\ \mathcal{A}_H\ e^{z\,\alpha_+}+ \mathcal{B}_H\ 
e^{z\,v_\omega/\overline D}
\end{equation}
where
\begin{align}
\label{BH}
\mathcal{B}_H=& \mathcal{A}_0+\, \mathcal{A}_h\frac{F}{F+G}\left\{
\frac{D_q\alpha_+ -v_\omega}{\overline D}
\left[\frac{1}{\lambda_+}+\frac{1}{\alpha_+ -v_\omega/\overline D}\right]
\right.\nonumber\\
+&\left.\ \frac{1}{\overline D \lambda_+}\left[
v_\omega-D_q(\alpha_+ +\lambda_+)+
\frac{\lambda_+ -\alpha_-}{\lambda_+ -\beta_+}\
\frac{F\lambda_+ (-v_\omega+D_q \lambda_+)-G\,\Gamma_m/k_H}
{F(\lambda_+-\beta_-)}
\right]\right\}\nonumber\\
-&\ \frac{1}{\overline D \lambda_+}\ \mathcal{A}_{\lambda}\ \frac{1}{F+G}
\ \frac{\alpha_+ -\beta_-}{(\lambda_+ -\beta_+)(\lambda_+ -\beta_-)}\
\left[F\lambda_+ (-v_\omega+D_q \lambda_+)-G\Gamma_m/k_H \right]
\end{align}
and
\begin{equation}
\label{AH}
\mathcal{A}_H=-\ \mathcal{A}_h\ \frac{F}{F+G}\
\frac{D_q\, \alpha_+ -v_\omega}{\overline D \alpha_+ -v_\omega}
\end{equation}
with
\begin{align}
\label{Alambda}
\mathcal{A}_\lambda=& \frac{2}{\sqrt{v_\omega^2+4\Gamma_1 D_h}
+\sqrt{v_\omega^2+4\Gamma_2 D_h}}
\int_0^\infty d\zeta\ \tilde{\gamma}_{\widetilde{H}_{-}}
(\zeta)\,e^{-\zeta \lambda_+}
\nonumber\\
\mathcal{A}_0=& \frac{1}{\overline D \lambda_+}\
\int_0^\infty d\zeta\ f_+(z)\ e^{-\zeta \lambda_+}
\end{align}
and
\begin{equation}
\label{lambda}
\lambda_\pm= \frac{1}{2\overline D}\left\{v_\omega\pm
\sqrt{v_\omega^2+4 \overline \Gamma\ \overline D} \right\}\ .
\end{equation}

Since we assume the sphalerons are inactive inside the bubbles, the
baryon density is constant in the broken phase and satisfies,  
in the symmetric phase, an equation where $n_L$ acts as a 
source~\cite{hn} and there is an explicit sphaleron-induced relaxation 
term~\cite{Shapo,plus}
\begin{equation}
\label{ecbaryon}
v_\omega n'_B(z)=-\theta(-z)\left[n_F \Gamma_{ws} n_L(z)
+ \mathcal{R}n_B(z)\right]
\end{equation}
where $n_F=3$ is the number of families and $\mathcal{R}$ is the relaxation
coefficient~\cite{Shapo},
\begin{equation}
\label{rel}
\mathcal{R}=\,\frac{5}{4}\, n_F\, \Gamma_{ws}\ .
\end{equation}
Eq.~(\ref{ecbaryon}) can be solved analytically and gives, in the broken
phase $z\ge 0$, a constant baryon asymmetry,
\begin{equation}
\label{nBsol}
n_B=-\,\frac{n_F \Gamma_{ws}}{v_\omega} \int_{-\infty}^0
dz\, n_L(z)\ e^{z\mathcal{R}/v_\omega}\ .
\end{equation}

Using now the explicit solutions for $n_H$ and $n_h$ given in 
Eqs.~(\ref{nHsol}) and (\ref{nhsol}), we can cast the explicit solution for
the baryon asymmetry as,
\begin{equation}
\label{nBfin}
n_B=\, n_F\, \Gamma_{ws}\ \frac{5 k_Q k_B+8 k_T k_B-9k_Q k_T}
{k_H \left( k_B+9 k_Q+9 k_T \right)}
\left\{\frac{\mathcal{A}_H + \mathcal{A}_h}{\mathcal{R}+v_\omega \alpha_+}
+\frac{\overline D \mathcal{B}_H}{\overline D \mathcal{R}+v_\omega^2}
\right\}
\end{equation}
where all symbols used in Eq.~(\ref{nBfin}) have been previously defined. 

The validity of our analytical approximation is guaranteed by the
dominance of $n_H$
over $n_h$, which in turn is related to
the $\tan\beta$ suppression of $\tilde{\gamma}_{\widetilde{H}_{-}}$
and the presence of $\Gamma_\mu$. In fact were we working in the limit
$\Gamma_\mu\to\infty$ we would find that the density $n_h$ is negligible. 
On the other hand, in the limit 
$\Gamma_\mu\to 0$ and $\tan\beta\simeq 1$ we would really
expect $n_h> n_H$, due to the dominance of 
$\tilde{\gamma}_{\widetilde{H}_{-}}$ over $\tilde{\gamma}_{\widetilde{H}_{+}}$,
at least for large values of $m_A$ where the $\Delta\beta$ suppression of 
$\tilde{\gamma}_{\widetilde{H}_{+}}$ is more severe. However 
small values of $\tan\beta$,
as we noticed earlier in this paper, are strongly disfavored 
in our scenario by recent LEP bounds on the Higgs mass.
Hence, we have found that the
analytical approximation is accurate with an error which depends on the
chosen values of the supersymmetric parameters, but it is always much
smaller than the other uncertainties involved in the final calculation. In
section~\ref{numerical} we will provide explicit comparison with the
numerical result, while all plots will be done using the numerical solution
of system~(\ref{sistema}).

\section{Numerical results}
\label{numerical}
In this section we present the numerical results for the baryon asymmetry
computed in section~\ref{bau} and, in particular, of the baryon-to-entropy
ratio $\eta\equiv n_B/s$, where the entropy density is given by,
\begin{equation}
\label{entropia}
s=\frac{2\pi^2}{45}g_{eff}\, T^3
\end{equation}
with
$g_{eff}$ being the effective number of relativistic degrees of freedom.
The profiles $H(z)$, $\beta(z)$ have been accurately computed in the 
literature~\cite{MOQ,last}. For the sake
of simplicity, in this paper we will use a kink
approximation~\cite{CQRVW} 
\begin{align}
\label{profiles}
H(z)= & \frac{1}{2}\, v(T)\,\left(1-\tanh\left[\alpha\left(1-
\frac{2\, z}{L_\omega}\right)\right]\right)\nonumber\\ & \nonumber\\
\beta(z)= &\beta-\frac{1}{2}\, \Delta\beta\,\left(1+\tanh\left[\alpha\left(1-
\frac{2\, z}{L_\omega}\right)\right]\right)\ .
\end{align}
This approximation has been checked
to reproduce the exact calculation of the Higgs profiles within a few percent
accuracy~\cite{mqs}, 
provided that we borrow from the exact calculation the values of the
thickness $L_\omega/2\alpha$ and the variation of the angle 
$\beta(z)$ along the
bubble wall, $\Delta\beta$, as we will do. In particular we will take
$\alpha\simeq 3/2$, $L_\omega=20/T$, and we have checked that 
the result varies
only very slowly with those parameters,
while we are taking the values of $\Delta\beta$ which are obtained
from the two-loop effective potential used in 
our calculation.
%

\begin{figure}[htb]
\vspace{.75cm}
\centering
\epsfig{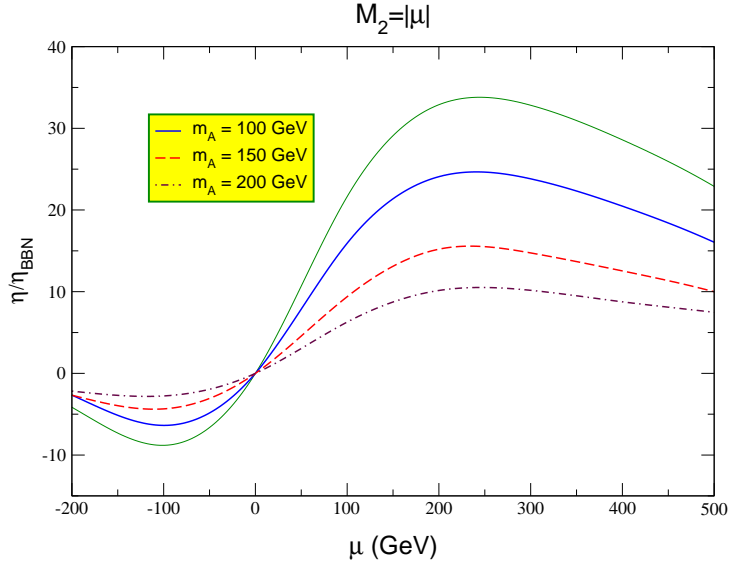}
\caption{Plot of $\eta/\eta_{\rm BBN}$ as a function of $\mu$ for $M_2=|\mu|$,
$m_A=100$ GeV (thick solid curve), $m_A=150$ GeV (dashed curve) and 
$m_A=200$ GeV (dash-dotted curve), and the rest of parameters as
indicated in the text. The thin solid curve corresponds to the case 
$m_A=100$ GeV when the approximate analytical solution in (\ref{nBfin})
is used.}     
\label{figm2mu}
\end{figure}
The calculation of the wall velocity $v_\omega$
is a very complicated phenomenon involving
the hydrodynamics of the bubble interacting with the 
surrounding plasma. Some progress has
been recently reported in this direction~\cite{vw} indicating that, in 
the case of the MSSM, the wall is extremely non-relativistic, 
$v_\omega\ll 1$, and can be as slow as $v_\omega=0.01$. 
Unless explicitly stated, 
in the numerical analysis of this section, 
we adopt the 
value $v_\omega=0.05$, although the variation of the baryon asymmetry 
with respect to $v_\omega$ will also be analyzed. 
The widths, $\Gamma_m$, $\Gamma_h$ and $\Gamma_Y$ are
as in Refs.~\cite{hn,CQRVW}, while we are taking 
$\Gamma_\mu\simeq 0.1\, T$ and $\rho= 1$, in agreement with the large
value we use for the left-handed third-generation squark masses, 
$m_Q\simgt 1$ TeV, 
which makes them decoupling from the thermal bath. On the other
hand, and consistently with the latter assumption (which is required 
to render the MSSM in agreement with the Higgs mass bounds
coming from LEP), the contribution to
$n_B$ from the squark source, $\tilde{\gamma}_Q$, is negligible.
The ``observable'' value for $\eta$ consistent with 
Big Bang Nucleosynthesis (BBN) has been considered to be  
$\eta_{\rm BBN}\sim 4\times 10^{-11}$~\cite{BBN}.
Finally we will consider the third generation squark 
mass and mixing parameters, $m_Q=1.5$ TeV and $A_t=0.5$ TeV,
and $\tan\beta=20$ and have checked that, for all 
plots in this section, the phase transition is strong enough first order,
$v(T_c)/T_c\simgt 1$, and the Higgs mass is, within the accuracy of
our calculations,  $m_h \simeq 110$--115 GeV.
These values are in rough agreement with present 95 \% C.L. bounds on the
Higgs mass coming from LEP, or even with 
the present excess of events observed at LEP, consistent with 
the detection of a SM-like Higgs at the runs with the highest
center of mass energies, $\sqrt{s}> 206$ GeV.
We will comment more about the LEP constraints in the next section.

In Fig.~\ref{figm2mu} we plot the ratio $\eta/\eta_{\rm BBN}$ for the 
values of the supersymmetric parameters that have just been described, 
$M_2=|\mu|$, $\sin\varphi_\mu = 1$
and several values of the pseudoscalar Higgs mass $m_A$. 
Therefore, 
since $\eta$ is (almost) linear in $\sin\varphi_\mu$, one
can read from Fig.~\ref{figm2mu} the value of $1/\sin\varphi_\mu$ that
would reproduce $\eta_{\rm BBN}$. This observation applies to all plots
presented in this section, where we have fixed $\sin\varphi_\mu=1$. 
It follows that the region of parameters where we find
$|\eta/\eta_{\rm BBN}|<1$, is forbidden in all plots.
For the value $m_A=100$ GeV, we have presented both the
exact result (thick solid curve), based on the numerical solution of 
Eqs.~(\ref{sistema}), and the approximate result (thin solid curve), based on
the approximate analytical 
solution (\ref{nBfin}). We see that for values where
$n_B/s$ is sizeable the discrepancy between the analytical and the numerical
result is $\simlt$ 30 \%. 
For the other curves in Fig.~\ref{figm2mu}, as well as
for the rest of plots in this paper, 
we will use the (exact) numerical solution of Eqs.~(\ref{sistema}).
%

\begin{figure}[htb]
\vspace{.75cm}
\centering
\epsfig{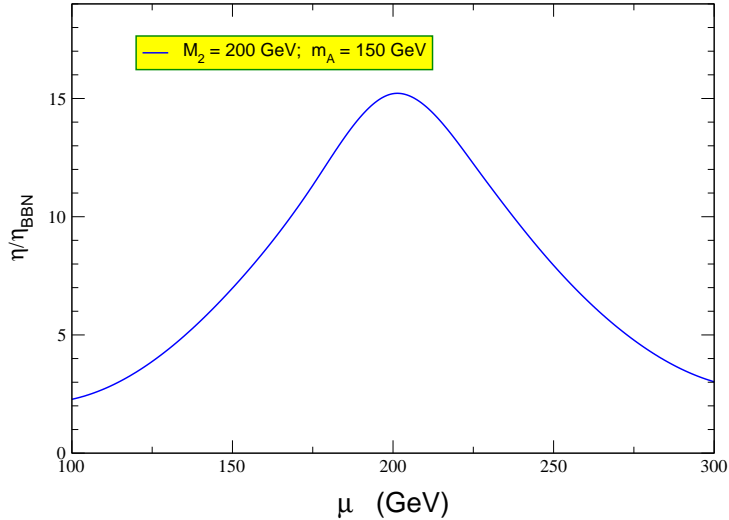}
\caption{Plot of $\eta/\eta_{\rm BBN}$ as a function of 
$\mu$ for $M_2=200$ GeV and $m_A=150$ GeV.}     
\label{figmu}
\end{figure}
We are, in Fig.~\ref{figm2mu}, close to the resonance region
discussed in Ref.~\cite{CQRVW}, which is smoothed
by the all order resummation in Higgs mass insertions. The departure from the
resonance is exemplified in Fig.~\ref{figmu}, where we plot 
$\eta/\eta_{\rm BBN}$ as a function of $\mu$ for $M_2=200$ GeV, 
$m_A=150$ GeV and the other supersymmetric parameters as in 
Fig.~\ref{figm2mu}.

\begin{figure}[htb]
\vspace{.75cm}
\centering
\epsfig{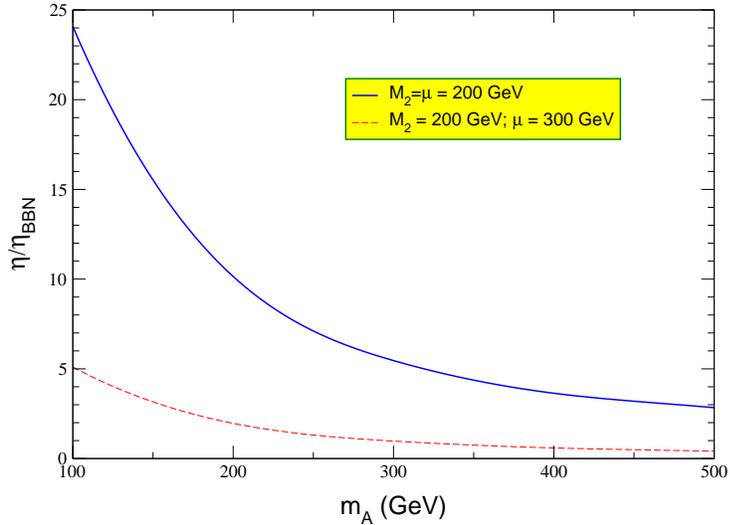}
\caption{Plot of $\eta/\eta_{\rm BBN}$
as a function of $m_A$ for $M_2=\mu=200$ GeV (solid curve)
and $M_2=200,\,\mu=300$ GeV (dashed curve).}  
\label{figma}
\end{figure}

\begin{figure}[htb]
\vspace{.75cm}
\centering
\epsfig{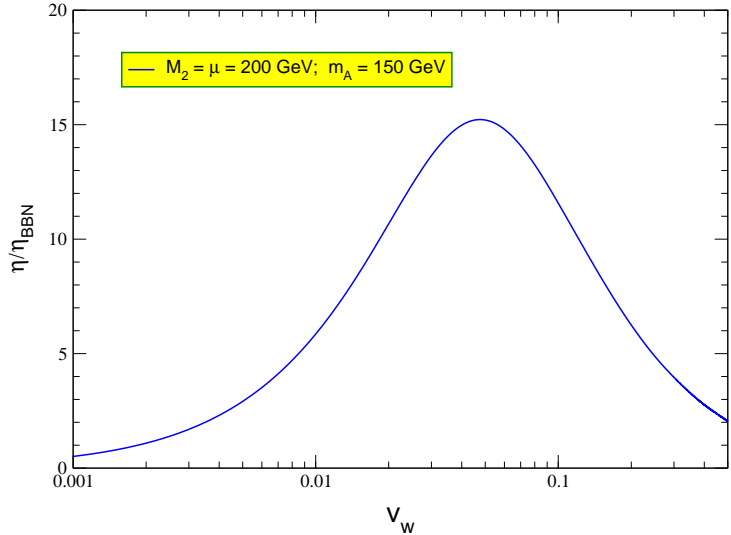}
\caption{Plot of $\eta/\eta_{\rm BBN}$ as a function of
$v_w$ for $M_2=\mu=200$ GeV and $m_A=150$ GeV.}     
\label{figvw}
\end{figure}

In Fig.~\ref{figma} we plot $\eta/\eta_{\rm BBN}$
as a function of $m_A$ for $M_2=\mu=200$ GeV (solid curve) and 
$M_2=200$ GeV,
$\mu=300$ GeV (dashed curve), and other supersymmetric parameters as in 
Fig.~\ref{figm2mu}.
Finally in Fig.~\ref{figvw} we plot $\eta/\eta_{\rm BBN}$ as a function of
$v_w$ for $M_2=\mu=200$ GeV, $m_A=150$ GeV and the other parameters as in 
Fig.~\ref{figm2mu}. The maximum of this curve comes from the interplay between
the relaxation and source terms in the equation for $n_B$, 
Eq.~(\ref{ecbaryon}).

%

The numerical results exhibited in the plots of this section are an 
improvement of our previous results, Ref.~\cite{CQRVW}, and include the
all order resummation of the Higgs mass insertions in the current
determination, as well as inclusion of finite $\Gamma_\mu$-effects in the
diffusion equations. Since the first of these effects smooths out the
resonant behaviour, which enhances the determination of $n_B$ for 
$M_2=|\mu|$, while the second one slightly enhances $n_B$, our present
numerical results are in rough agreement with those of Ref.~\cite{CQRVW}.
On the other hand if we compare our numerical results with the recent ones
of Ref.~\cite{plus}, that use WKB methods  and values of
$\rho < 1$ to deduce the source terms in the
diffusion equations, we observe a discrepancy of a few orders of
magnitude. However, we have been communicated~\cite{Kimmo} by the authors
of Ref.~\cite{plus} to have detected a problem in their numerical codes which
enhances their numerical results by some orders of magnitude and that might 
explain part of this discrepancy. As explained above, large values of
$m_Q$, implying $\rho = 1$, are necessary in order to fulfill the present
experimental Higgs mass bounds.

\section{Higgs mass constraints}
\label{higgs}

   In this section, we shall comment on the constraints coming from
Higgs searches at LEP. The LEP experiments at CERN have
collected data during the year 1999 at various energies
between 192 GeV and 202 GeV, for a total integrated luminosity
of about 900 $pb^{-1}$. A combined limit on the Standard Model Higgs
mass of about 108 GeV at the 95 $\%$ C.L. was obtained, due
to the absence of any significant 
Higgs signal in the LEP data~\cite{lephiggs}. 
Preliminary results
of this year run~\cite{lepchiggs} show that this limit
moved up by a few GeV (up to about 113.2 GeV). More interesting,
a slight excess of events, about 3 standard deviations above
the SM predictions, has been observed, consistent with 
a SM like Higgs in the range of masses of about 113--116 GeV. 

   The present Higgs mass constraints become particularly relevant for small
values of $\tan\beta$, $\tan\beta <$~5. In this case, due to the
behaviour of the Higgs boson couplings to fermions and
gauge bosons, the SM Higgs mass constraints translate 
with almost no variations into a bound on the lightest CP-even
Higgs boson mass~\footnote{In the presence of CP-violation,
the Higgs mass eigenstates will not be CP-eigenstates. In our analysis
we have used the CP conserving structure for the Higgs sector. This 
should lead to a good approximation if 
CP-violating effects in the Higgs potential are small,
as happens when arg$(\mu_c A_t) \simeq 0$. 
A more general analysis,
similar to the one performed in Refs.~\cite{CEPW,CEPW2} 
would be appropriate to consider more general CP-violating effects.}. 
For values of $v(T_c)/T_c \simgt 1$ and $\tan\beta < 5$, 
and for left-handed stop masses smaller than $\sim$ 3 TeV, 
the lightest CP-even 
Higgs mass never exceeds 105 GeV. Therefore, the mechanism
of electroweak baryogenesis demands either values of $\tan\beta >5$
or unnaturally large values of $m_Q$~\footnote{
We have checked that, for the values of
the stop mixing
parameters consistent with electroweak baryogenesis, no significant
modification of these bounds is obtained after considering CP-violating
effects in the Higgs potential~\cite{CEPW}.}.

Large values of $\tan\beta$ move
the value of the  Higgs boson mass, with relevant
couplings to the gauge bosons, to larger values.
However, if the values of the left-handed stop parameters
are restricted to be below 3 TeV, for $v(T_c)/T_c \simgt 1$, the
Higgs mass cannot exceed 115 GeV. Observe that these
values are a few GeV higher than those obtained previously
in Ref.~\cite{CQW2}, since in that reference we restricted ourselves
to the case of left-handed stop masses below 1 TeV. The observed 
excess of events, with $b\bar{b}$ invariant 
masses of about 114 GeV, would be consistent
with electroweak baryogenesis for large values of $\tan\beta$ and
large values of the left-handed stop mass parameters $m_Q \simgt 1 $ TeV,
as the ones considered in the previous section.

What would happen if the excess of events present at LEP would not
correspond to a Higgs signal, but would turn out to be a statistical
fluctuation with the final outcome of an ultimate exclusion limit
for a SM-like Higgs with mass below 115 GeV? 
To analyze this, let us stress that
at large values of $\tan\beta$
the coupling of this Higgs boson to bottom quarks 
can be  significantly lower than in the SM~\cite{CMW,CEPW}
with a corresponding reduction of the Higgs mass bound.
These variations can only occur for small values of the CP-odd
Higgs mass $m_A$, of order of the lightest CP-even 
Higgs boson mass. Unlike the case of
small values of $\tan\beta$, 
for values of $\tan\beta > 10$,
the values of $v(T_c)/T_c$ are only weakly dependent
on the exact value of $m_A$. Intuitively, this can be understood by
the fact that for large values of $\tan\beta$, the CP-odd Higgs
can be approximately identified with the imaginary part of the 
neutral component of the Higgs doublet $H_1$, while the Higgs
doublet which acquires vacuum expectation value is mainly
$H_2$ ($v_2 \gg v_1)$.

For the values of $A_t$ and $\mu$ consistent with electroweak baryogenesis,
a reduction of the coupling of the CP-even Higgs boson to
the bottom quark would demand not only small values
of $m_A \simeq 100$--150 GeV, but also large values
of $\tan\beta > 10$ and of $|\mu A_t|/m_Q^2 > 0.1$
(the larger $\tan\beta$, the easier 
suppressed values of the bottom quark coupling are obtained).
We have checked that, assuming small CP-violating effects
in the Higgs potential and in the Higgs-fermion couplings,
and for values of $m_Q \simeq 1$ TeV, $v(T_c)/T_c \simgt 1$ 
and $|\mu| < 500$ GeV, a significant reduction of the coupling
of the Higgs to bottom quarks only occur for $\tan\beta \simgt 30$.
Therefore, if the excess of events observed at LEP is not
associated with a Higgs signal,
strong constraints on the electroweak baryogenesis scenario within
the MSSM will be obtained. 

\section{Conclusions}
\label{conclusion}

      In this article, we have performed a computation of the
scalar- and fermion- CP-violating currents induced by the 
expansion of a true-vacuum bubble in the false vacuum plasma,
within the framework of the minimal supersymmetric standard
model. We made use of the Keldysh formalism and we have
defined a systematic way of obtaining the currents in an
expansion of derivatives of the Higgs fields, to all orders
of the Higgs background insertions. 

      Although our method is similar to the one 
previously used by some of us in Ref.~\cite{CQRVW},
our results differ from those presented in our previous work 
in several respects.
First of all, they include a resummation of corrections
associated with higher order of the Higgs background
insertions. These corrections have two important
effects. The first one is to 
dilute the resonant behaviour obtained in Ref.~\cite{CQRVW}
for values of $|\mu| = M_2$. The second one is the 
appearance of a contribution proportional to $H_2\partial^{\mu} H_1
+ H_1 \partial^{\mu} H_2$ to the vector Higgsino current
$j_{\hS_{+}}^\mu(z)$. This
means that, as first observed in Refs.~\cite{hn,nuria}, the
vector Higgsino current does not vanish for large values
of the CP-odd Higgs mass. Our method provides a  
way of obtaining the value of this non-vanishing
contribution in a self-consistent way.
In addition, we have also computed 
the axial Higgsino current $j_{\hS_{-}}^\mu(z)$, whose
components are proportional to
$H_2 \partial^\mu H_1 + H_1 \partial^\mu H_2$.
Therefore, as first observed in Ref.~\cite{plus}, 
the chiral current is not suppressed for large values
of the CP-odd Higgs mass and hence may become
relevant in this regime.

The vector and axial Higgsino currents, $j_{\hS_{\pm}}^\mu(z)$,
were used to determine
the baryon asymmetry of the Universe, $n_B/s$. The
computation of $n_B$ demands the solution of diffusion
equations, with sources determined through $j_{\hS_{\pm}}^\mu(z)$.
Following the method developed in Refs.~\cite{hn,Toni2}, 
we assumed that 
the sources are proportional 
to the temporal component of the currents, with a constant of 
proportionality given by the Higgsino width. Within this
approximation, we computed
the functional dependence of $n_B$ on the soft
supersymmetry breaking parameters and on the bubble wall 
parameters. The most important parameters turn out to 
be the gaugino and Higgsino mass parameters, $|\mu_c|$ and
$M_2$, their relative phase arg$(\mu_c M_2)$, (equal to $\varphi_{\mu}$
in the basis in which $M_2$ is real) as well as
the CP-odd Higgs mass $m_A$ and $\tan\beta$. We have also
required that the condition of preservation of the baryon
asymmetry $v(T_c)/T_c \simgt 1$ is fulfilled, what
demands a light right-handed
stop and, due to the present Higgs mass
constraints coming from LEP (see section \ref{higgs}), also large
values of the ratio of Higgs vacuum expectation values,
$\tan\beta > 5$.

   Under the above conditions, we have determined the value
of $n_B$, compared to the value predicted by Big Bang 
Nucleosynthesis, for a value of $\sin \varphi_{\mu} =$ 1. The
ratio of the theoretically obtained to the BBN 
predicted baryon asymmetry
can be reinterpreted as the inverse of the value of 
$\sin\varphi_{\mu}$ needed to obtain a value of $n_B$ 
in agreement with the
BBN predictions. We conclude that,
for small values of $m_A \simeq 100$ GeV and $|\mu| \simeq M_2$,
values as low as $\varphi_{\mu} \simeq 0.04$ can lead to
acceptable values of $n_B$. The predicted
value of the phase $\varphi_{\mu}$
increases for larger
values of $m_A$ and/or for $|\mu| \neq M_2$, but still there
is a large fraction of parameter space in which the computed
baryon number is in good agreement with BBN predictions, for
phases such that $\sin\varphi_{\mu} \simeq 0.04$--1. 

Values of 
$\varphi_{\mu} \simgt 0.04$ can  lead to acceptable phenomenology
if either peculiar cancellations in the squark and slepton 
contributions to the neutron and electron electric dipole moments (EDM) 
occur~\cite{cancelacion}, and/or if the first 
and second generation of squarks are heavy~\cite{Alex}. This second
possibility is quite appealing and, as has been recently
demonstrated~\cite{DarkM}, 
leads to acceptable phenomenology, including
the dark matter constraints~\footnote{Third generation squarks would still
contribute to the neutron and electron EDM, 
via two loop diagrams
involving the would-be CP-odd Higgs boson~\cite{CKP}. These
contributions can become sizeable at large values of $\tan\beta$,
although they tend to be suppressed for small values of the mixing in
the stop sector, as the ones required for electroweak baryogenesis.}.

Another important observable which, similarly to the
value of the baryon number, depends on the precise
value of the mass parameters in the gaugino, Higgsino and
third generation squark sectors, as well as on the charged Higgs mass,
is the rate of the rare decay $b \rightarrow s \gamma$~\cite{BBMR}. 
For small values of the charged Higgs and stop masses, and
for moderate values of $A_t/m_Q$ and $|\mu|/m_Q$,
the chargino-stop contribution, as well as the charged
Higgs contribution, may become large for large values of
$\tan\beta$~\cite{Riccardo,Gambino,CGNW}. 
In scenarios with heavy first and second generation squarks,
however, flavor violation couplings involving the third
generation squarks could be non-negligible~\cite{Alex} and therefore
the gluino-sbottom contributions to this rare decay rate may be 
enhanced~\cite{Borzumati}.
Since these last contributions are strongly model dependent,
and may be larger than the charged Higgs and chargino-stop
ones, we have not imposed the $b \rightarrow s \gamma$
constraints in our analysis.

  Finally, we have discussed the effect of the Higgs mass constraints 
coming from LEP. The preliminary data coming from the LEP experiments
imply a lower bound on the mass of a SM-like Higgs boson of about 113 GeV. A
small excess, consistent with a SM-like Higgs boson with a mass
slightly above that value has also been observed. These relatively large
values of the Higgs mass are consistent with electroweak baryogenesis
within the MSSM if the value of $\tan\beta$ is large, $\tan\beta >5$,
if the  left-handed stops are heavy $m_Q \simgt 1$ TeV, and if the stop
mixing parameter is not small, $A_t \simgt 0.25 \; m_Q$. On the other hand,
for these values of the Higgs mass, values of $A_t \simgt 0.4 \; m_Q$ make
the phase transition weaker, leading to values of $v(T_c)/T_c$ that are in
conflict with the condition of preservation of the baryon asymmetry.
It is important to emphasize, however, that
even if the CP-even Higgs
boson coupling to the gauge boson is SM like, it can
evade the LEP bounds if its coupling to the bottom quark is strongly
suppressed, what can occur for very large values of $\tan\beta$,
$\tan\beta \simgt 30$. More relevantly, if the excess of events
at LEP has its origins in the presence of a SM-like Higgs boson
of mass of about 113--115 GeV, one of the predictions of electroweak
baryogenesis, namely the presence of a light neutral Higgs boson with
SM-like couplings to the gauge bosons and a 
mass not larger than 115 GeV would have been fulfilled.

\section*{Acknowledgements}

M.Q. would like to thank the Laboratoire de Physique Th\'eorique,
Ecole Normale Sup\'erieure (Paris), where part of this work has been done,
and Kimmo Kainulainen for discussions and correspondence concerning
Ref.~\cite{plus}. M.C. and C.E.M.W. would like to thank the Aspen Center
for Physics and the Physics Department of the University of California,
Santa Cruz, where part of this work has been done.  This work 
has been supported
in part by the US Department of Energy, High Energy Physics Division,
under Contracts DE-AC02-76CHO3000 and W-31-109-Eng-38, 
by CICYT, Spain, under contract AEN98-0816, and by EU under TMR contract
ERBFMRX-CT96-0045 and RTN contract HPRN-CT-2000-00152.

\end{document}